\documentclass[prl,aps,twocolumn,superscriptaddress]{revtex4}
\usepackage{amsmath,graphicx,epsfig}

\usepackage{epstopdf}
\usepackage{euscript}
\usepackage{amsfonts}
\usepackage{amssymb}
\usepackage{float}
\usepackage{url}

\def\prn#1{{\left(#1\right)}}

\def\abrk#1{{\langle#1\rangle}}

\def\bra#1{{\langle#1|}}

\def\cg(#1,#2)(#3,#4)(#5,#6){\bra{#1,#2,#3,#4}#5,#6\rangle}

\def\ts#1{{_{\mbox{\scriptsize #1}}}}

\def\threej(#1,#2)(#3,#4)(#5,#6){\begin{pmatrix}#1&#3&#5\\#2&#4&#6\end{pmatrix}}
\def\sixj(#1,#2,#3)(#4,#5,#6){\begin{Bmatrix}#1&#2&#3\\#4&#5&#6\end{Bmatrix}}
\def\ninej(#1,#2,#3)(#4,#5,#6)(#7,#8,#9){\begin{Bmatrix}#1&#2&#3\\#4&#5&#6\\#7&#8&#9\end{Bmatrix}}

\def\mb{\mathbf}

\newlength{\defbaselineskip}
\newcommand{\setlinespacing}[1]%
           {\setlength{\baselineskip}{#1 \defbaselineskip}}

\begin{document}

\title{Precessing Ferromagnetic Needle Magnetometer} 

\date{\today}



\author{Derek F. Jackson Kimball}
\affiliation{Department of Physics, California State University -- East Bay, Hayward, California 94542-3084, USA}

\author{Alexander O. Sushkov}
\affiliation{Department of Physics, Boston University, Boston, Massachusetts 02215, USA}

\author{Dmitry Budker}
\affiliation{Helmholtz Institute Mainz, Johannes Gutenberg University, 55099 Mainz, Germany}
\affiliation{Department of Physics, University of California at Berkeley, Berkeley, California 94720-7300, USA}
\affiliation{Nuclear Science Division, Lawrence Berkeley National Laboratory, Berkeley, California 94720, USA}

\begin{abstract}
A ferromagnetic needle is predicted to precess about the magnetic field axis at a Larmor frequency $\Omega$ under conditions where its intrinsic spin dominates over its rotational angular momentum, $N\hbar \gg I\Omega$ ($I$ is the moment of inertia of the needle about the precession axis and $N$ is the number of polarized spins in the needle). In this regime the needle behaves as a gyroscope with spin $N\hbar$ maintained along the easy axis of the needle by the crystalline and shape anisotropy. A precessing ferromagnetic needle is a correlated system of $N$ spins which can be used to measure magnetic fields for long times. In principle, by taking advantage of rapid averaging of quantum uncertainty, the sensitivity of a precessing needle magnetometer can far surpass that of magnetometers based on spin precession of atoms in the gas phase. Under conditions where noise from coupling to the environment is subdominant, the scaling with measurement time $t$ of the quantum- and detection-limited magnetometric sensitivity is $t^{-3/2}$. The phenomenon of ferromagnetic needle precession may be of particular interest for precision measurements testing fundamental physics.
\end{abstract}



\maketitle

For an ensemble of $N$ independent particles prepared in a coherent superposition of quantum states, the standard quantum limit (SQL) on the precision of a measurement of the phase $\phi$ is given by \cite{Bra75}
\begin{align}
\Delta \phi \approx \sqrt{\frac{\Gamma\ts{rel}t}{N}}
\label{Eq:SQL-for-phase}
\end{align}
after time $t \gg 1/\Gamma\ts{rel}$, where $\Gamma\ts{rel}$ is the relaxation rate of the coherence. Equation~\eqref{Eq:SQL-for-phase} represents a random walk in phase with step size $1/\sqrt{N}$ consisting of $\Gamma\ts{rel}t$ steps. In cases where the goal is to measure a frequency $\Omega = \phi/t$, there is an analogous SQL on the precision of a frequency measurement,
\begin{align}
\Delta \Omega \approx \sqrt{\frac{\Gamma\ts{rel}}{Nt}}~.
\label{Eq:SQL-for-frequency}
\end{align}
For a measurement subject to the SQL, the minimum possible measurement uncertainty is obtained when $\Gamma\ts{rel}$ is made as small as possible. In the limit where $\Gamma\ts{rel} \rightarrow 0$, the precision becomes constrained by the duration of the measurement, so in Eqs.~\eqref{Eq:SQL-for-phase} and \eqref{Eq:SQL-for-frequency}, $\Gamma\ts{rel}$ is replaced by $1/t$.

However, if the particles' time evolution is correlated, the SQL can be circumvented for times shorter than the coherence time ($1/\Gamma\ts{rel}$) \cite{Win92,Hue97,Auz04}.  Extensive experimental efforts involving quantum entanglement, squeezed states, and quantum nondemolition (QND) measurement strategies have been made to take advantage of this potential improvement in measurement sensitivity \cite{Sha10,Was10,Sew12,Wol12}. In this Letter we draw attention to a system which can, in principle, surpass the SQL on measurement of spin precession in a different way: by rapid averaging of quantum uncertainty.

In particular, we consider the measurement of magnetic fields. The most precise magnetic field measurements are based on the techniques of optical atomic magnetometry \cite{Bud07,Bud13book}: $N$ atomic spins are optically polarized and their precession in a magnetic field $B$ is measured using optical rotation of probe light \cite{Bud02review}. Depending on its magnitude, the value of $B$ is either extracted from measurement of the Larmor frequency $\Omega = g \mu_B B / \hbar$ or the accrued spin precession angle $\phi = \Omega t$ if $\phi \ll 1$ during the measurement time $t$ ($g$ is the Land\'e g-factor and $\mu_B$ is the Bohr magneton). Optical atomic magnetometers with paramagnetic atoms have achieved sensitivities $\delta B \approx 10^{-12}~{\rm G/\sqrt{Hz}}$ \cite{Was10,Dan10,She13}, close to the SQL of $\delta B \approx 10^{-13}~{\rm G/\sqrt{Hz}}$ \cite{All02,Kom03,Sav05}.

Remarkably, there is a system that can be used for magnetometry that, in principle, can surpass the SQL on measurement of spin precession: a ferromagnetic particle, for example in the shape of a needle. In fact, such a device is reminiscent of the very first magnetometer developed by Gauss in the 1830s -- the ``Unifilarmagnetometer'' -- a ferromagnetic needle suspended from a gold fiber \cite{Gau1833}. A new class of ferromagnetic-needle magnetometers is possible based on the observation that for sufficiently small torques a magnetic needle will precess about the field axis at the Larmor frequency \cite{Bud08book} instead of orienting itself along the field direction (or oscillating about the field direction in the case of an underdamped system). These two regimes of behavior can be understood in analogy with the behavior of a gyroscope in a gravitational field: as long as the angular momentum along the spin axis is sufficiently large, the gyroscope precesses about the direction of the gravitational field; if the angular momentum along the spin axis dips below a threshold value, the gyroscope tips over. The latter tipping behavior is analogous to the usual behavior of a ferromagnetic needle in a magnetic field (e.g., a compass needle and the concept of Gauss's Unifilarmagnetometer), while the former behavior is analogous to the precession of an isolated atomic spin in a magnetic field. In the case of the ferromagnetic needle magnetometer proposed here, it is the collective intrinsic spin of the needle that provides the angular momentum along the axis of the needle: the needle will precess as long as the needle's intrinsic spin angular momentum exceeds its rotational angular momentum.

\begin{figure}
\center
\includegraphics[width=2.5 in]{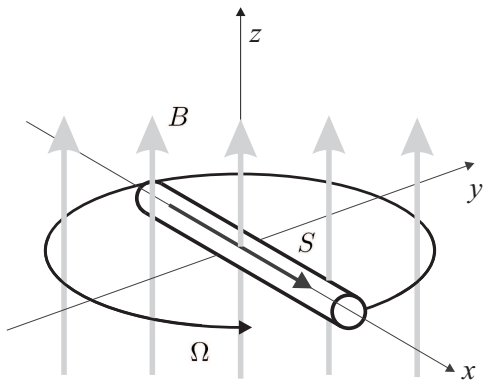}
\caption{A ferromagnetic needle with spin $S = N\hbar$ along its long axis precesses at a frequency $\Omega$ in a magnetic field $B \ll B^*$ [Eq.~\eqref{Eq:B-star}].}\label{Fig:PrecessingNeedle}
\end{figure}

Consider a freely floating, cylindrically shaped needle of length $\ell$, radius $r$, and mass $M$ made of a ferromagnetic material such as cobalt (Fig.~\ref{Fig:PrecessingNeedle}). We assume that the needle's axis is along the crystalline anisotropy axis, and that the needle is small enough to contain only a single magnetic domain, with the remanent magnetization along the needle's axis \cite{Aha88}; single-domain ferromagnetic cobalt needles with $\ell \approx 10~{\rm \mu m}$ and $r \approx 1~{\rm \mu m}$, for example, have been fabricated \cite{Sey01}. For such a single-domain ferromagnetic needle, the two equivalent ground states for the remanent magnetization along the needle's long axis are separated by a large energy barrier due to the exchange force and anisotropy field. Thus, under the conditions of near-zero magnetic field and cryogenic temperatures considered here, the collective spin will remain oriented along the needle's axis essentially forever \cite{Led94}. Furthermore, unlike in the case of a ferromagnet with multiple domains, in this case there is no magnetization noise from the motion of domain walls \cite{Chi97}.

In order to be in the regime where the needle precesses, the mechanical orbital angular momentum $L$ of the needle due to its precession, $L=I\Omega$ (where $I = M\ell^2/12$ is the needle's moment of inertia and $\Omega$ is the precession frequency), must be less than the angular momentum due to the intrinsic spin of the $N$ polarized electrons $S \approx N\hbar$, leading to the condition:
\begin{align}
\Omega \ll \Omega^* = \frac{N\hbar}{I} = \frac{12\hbar}{m_a \ell^2}~,
\label{Eq:Omega-star}
\end{align}
where $m_a$ is the atomic mass. In order for the needle to exhibit gyroscopic behavior, according to Eq.~\eqref{Eq:Omega-star} the background magnetic field $B$ must satisfy:
\begin{align}
B \ll B^* = \frac{\hbar \Omega^*}{g \mu_B}~.
\label{Eq:B-star}
\end{align}
For a single-domain needle with $\ell \approx 10~{\rm \mu m}$ and $r \approx 1~{\rm \mu m}$, $\Omega^* \approx 100~{\rm s}^{-1}$ and $B^* \approx 10^{-5}~{\rm G}$ (with $g \approx 1$ for cobalt), a field value that can be achieved in the laboratory with appropriate shielding \cite{Bud13book}. Depending on the application, different needle dimensions can be considered: for smaller needles the field requirements are relaxed since $B^*$ increases, but at the cost of sensitivity as discussed below.  The aspect ratio of the needle can also be optimized for the best magnetometer performance, with the caveat that depending on the aspect ratio the needle may transition into multi-domain behavior \cite{Aha88,Sey01}.

The dynamical process through which the intrinsic spin is coupled to the mechanical motion of the needle is described by the Landau-Lifshitz-Gilbert equation \cite{Lan35,Gil04}.  As a first approximation, we treat the needle as consisting of two coupled subsystems, the needle's crystal lattice, whose long axis is specified by the unit vector $\mb{a}$, and the collective spin $\mb{S}$ (the macrospin approximation \cite{Xia05,Bra08}). Furthermore at this initial stage, in order to simplify the discussion, we neglect noise related to external perturbations (e.g., collisions with gas molecules and black-body photons) and internal degrees of freedom [e.g., lattice vibrations (phonons), spin waves (magnons), and thermal electric currents].  In equilibrium, $\mb{S}$ is oriented along $\mb{a}$: $S_a = N\hbar$. If $\mb{S}$ is rotated with respect to $\mb{a}$, there is a torque exerted on the lattice. The lattice relaxes back to its equilibrium orientation where $\mb{a}$ is parallel to $\mb{S}$ at a rate $\Gamma_G \approx \alpha \omega_0$, where for bulk cobalt at room temperature the Gilbert constant is $\alpha \sim 0.01$ \cite{Bha74,Kat00,Sch95,Tse02} and the ferromagnetic resonance frequency is $\omega_0 = g \mu_B H\ts{eff}/\hbar$, where $H\ts{eff}$ is the effective internal magnetic field (anisotropy and exchange fields) acting on the spins. For bulk cobalt $\omega_0 \sim 10^{11}~{\rm s^{-1}}$ \cite{Wat61} and thus $\Gamma_G \sim 10^{9}~{\rm s^{-1}}$ \cite{Bha74,Gil07} ($\Gamma_G$ is even faster at low temperatures \cite{Bha74} and for micron-scale needles \cite{Sey01}). Under these conditions the system should relax to equilibrium ($\mb{S}$ along $\mb{a}$) with a characteristic time scale $\lesssim 1~{\rm ns}$.

The macroscopic dynamics of the needle can thus be understood as follows. Suppose that initially the needle is prepared as in Fig.~\ref{Fig:PrecessingNeedle}, at rest, and $B=0$. When a magnetic field $B \ll B^*$ is suddenly turned on, $\mb{S}$ experiences a torque and begins to precess at the Larmor frequency $\Omega$. The lattice, however, has inertia and undergoes angular acceleration due to the torque that arises when $\mb{S}$ is tilted with respect to $\mb{a}$. Since the relative motion between $\mb{S}$ and $\mb{a}$ is damped, $\mb{a}$ re-aligns with $\mb{S}$ after a time $\sim 1/\Gamma_G$. After this transient, the needle lattice rotates at frequency $\Omega$ with constant angular momentum $L = I\Omega$, and no further torque is exerted between the lattice and spin since their motion is synchronized. From another point of view, the needle is a rigid rotor characterized by the orientation of its axis $\mb{a}$, its rotational angular momentum $\mb{L}$, and its spin $\mb{S}$. The angular momenta add to the total angular momentum $\mb{J} = \mb{S} + \mb{L}$, but the needle is always in the regime where the spin angular momentum dominates: $S \gg L$. As noted above, the spin-lattice interaction essentially locks $\mb{S}$ and $\mb{a}$ together, thus the motion of the needle is dominated by the behavior of $\mb{S}$. For example, if the needle is prepared at rest in zero magnetic field and a non-adiabatic (faster than $1/\Gamma_G$) impulse imparts some rotational angular momentum $L \ll S$ to the needle misaligning $\mb{S}$ and $\mb{a}$, the angular momenta precess around $\mb{J} = \mb{S} + \mb{L}$ for a time $1/\Gamma_G$, but at longer times $\mb{S}$ and $\mb{a}$ again become oriented along $\mb{J} \approx \mb{S}$.

The electron spins of a precessing needle, being coupled to the crystal lattice, act collectively, as opposed to the spins in a polarized gas that act independently and can dephase \cite{Bud13book}. The situation is analogous to the M\"ossbauer effect \cite{Mos58} where the entire crystal lattice recoils from emission of a gamma ray. Needle precession also bears some relation to the Barnett \cite{Bar14,Bar35} and Einstein-de Hass \cite{Ein15} effects, where coupling between magnetization and macroscopic rotation is observed (see also Refs.~\cite{Alz67,Los16}). In terms of the measurement of an external magnetic field, the key differences between the ferromagnetic needle and a gas of paramagnetic atoms are the strong spin correlations present in the needle (where the spins are oriented along $\mb{a}$ due to the exchange and anisotropy forces) and the fast averaging and relaxation of spin components transverse to the needle's axis $\mb{a}$ (maintaining the strong coupling between $\mb{a}$ and $\mb{S}$).

As a specific realization of a magnetic field measurement, suppose the needle is prepared at rest with $\mb{a}$ and $\mb{S}$ pointing along $\hat{\mb{x}}$ and immersed in a constant external magnetic field $\mb{B}=B\hat{\mb{z}}$ of unknown magnitude (but with $B \ll B^*$), as in Fig.~\ref{Fig:PrecessingNeedle}. Some time $t$ after the needle is prepared in this way, the spin projection along $\hat{\mb{y}}$ ($S_y$) is measured, for example, by using a Superconducting QUantum Interference Device (SQUID) to detect the magnetic flux through a pick-up loop oriented to measure $S_y$. The needle precesses around the magnetic field at $\Omega$; to determine $B$, we measure $S_y$ and extract the value of $B$ from the precession angle
\begin{align}
\phi = \Omega t = g \mu_B B t/\hbar \approx \frac{S_y}{S_x} \approx \frac{S_y}{N\hbar}~,
\label{Eq:spin-angle-B}
\end{align}
assuming $\phi \ll 1$ (although, in the end, this is not essential for the estimate of sensitivity). It should be noted that in lieu of preparing the needle at rest, its orientation can be measured by the SQUID at $t=0$ since for determination of $B$ what matters is the change in $\phi$ or the precession frequency $\Omega$.

In order to estimate the uncertainty of such a measurement, let us begin by considering the precision with which the precession of the needle can be measured using a SQUID to detect the changing magnetic flux as the needle rotates. This technique resembles the recently developed hybrid SQUID-GMR (giant magnetoresistive) sensor of Ref.~\cite{Pan04}. Consider a dc SQUID detector with dimensions similar to the needle \cite{Aws88,Muc01}; this is similar to experimental setups used, for example, to read-out micromechanical resonators \cite{Use11} and detect magnetic particles \cite{Hub08}. Assuming a SQUID pick-up loop placed $\approx \ell \approx 10~{\rm \mu m}$ away from the tip of the needle of radius $\approx \ell \sin \theta_m \approx 8.2~{\rm \mu m}$, where $\theta_m \approx 54.74^\circ$ is the magic angle, chosen to optimize the flux capture, a changing magnetic flux of amplitude $\Phi \approx 10^{-4}~{\rm G \cdot cm^2}$ would be measured; SQUID systems employing flux-locked loops have demonstrated sufficient dynamic range to accommodate such a flux change \cite{Cla04v1}. The sensitivity of low-temperature SQUIDs to flux changes is $\delta \Phi \lesssim 10^{-13}~{\rm G \cdot cm^2 / \sqrt{ Hz }}$ \cite{Aws88,Use11,Hub08}, which limits the angular resolution of the needle measurement to $\delta \phi\ts{det} \approx \delta \Phi /\Phi \lesssim 10^{-9}~{\rm rad}/\sqrt{\rm Hz}$. This translates into a detection-limited uncertainty in determination of the magnetic field given by
\begin{align}
\Delta B\ts{det} \approx 10^{-16} \prn{t[{\rm s}]}^{-3/2}~{\rm G}~.
\label{Eq:magnetometry-SQUID-limit}
\end{align}
Note the $t^{-3/2}$ scaling of the measurement uncertainty, a result of the gyroscopic stability of the needle that prevents it from executing a random walk in angular position since $S \gg L$. Since $\Phi$ is proportional to $N$, as long as the needle remains single-domain and the pick-up loop geometry can be optimized, $\Delta B\ts{det}$ scales as $1/N$.

Special care must be taken to minimize the effect of any back-action field generated by current induced in the pick-up loop by the needle's precession. There are several schemes to eliminate such effects \cite{Tes90,Led05}, essentially involving active or passive feedback systems using additional coils to cancel the back-action field at the location of the sample, and so this is not a fundamental limitation. A related issue is back-action noise generated by the SQUID itself: here, too, there are several successful techniques for back-action evasion \cite{Poo10,Hat11} that yield back-action noise at the level of the magnetometric sensitivity given by Eq.~\eqref{Eq:magnetometry-SQUID-limit}.

Notably, at sufficiently long measurement times $t$, $\Delta B\ts{det}$ far surpasses the SQL for $N$ independent atomic spins even under conditions where $\Gamma\ts{rel} = 0$. In the $\Gamma\ts{rel} \rightarrow 0$ limit, $\Delta \Omega = g\mu_B \Delta B/\hbar = 1/( t \sqrt{N} )$. For $N \approx 3 \times 10^{12}$ spins, we obtain $\Delta B\ts{SQL} \approx 7 \times 10^{-14} \prn{t[{\rm s}]}^{-1}~{\rm G}$, and thus $\Delta B\ts{SQL} / \Delta B\ts{det} \sim 10^3 \sqrt{t[{\rm s}]}$.

To understand how a magnetometer based on a precessing ferromagnetic needle could surpass the SQL on measurement of spin precession, let us first consider spin projection noise for the case of an isolated spin $\mb{S}$; i.e., we neglect the spin-lattice interaction (this is analogous to a gas of paramagnetic atoms with collective spin $\mb{S}$ and $\Gamma\ts{rel} = 0$ as considered above). If the experiment measuring $\phi$ described above is repeated many times for an isolated spin $\mb{S}$, the spread of the results is governed by the uncertainty principle
\begin{align}
\Delta S_y \Delta S_z \geq \frac{\hbar}{2} \left| \abrk{ S_x } \right| \approx \frac{\hbar^2 N}{2}~,
\label{Eq:spin-uncertainty}
\end{align}
and so from Eq.~\eqref{Eq:spin-angle-B}, assuming $\Delta S_y \approx \Delta S_z$,
\begin{align}
\Delta \phi \approx \frac{1}{\sqrt{N}}~.
\label{Eq:spin-angle-uncertainty-no-lattice}
\end{align}
This is the well-known spin-projection noise \cite{Bud13book} that results in the SQL of Eqs.~\eqref{Eq:SQL-for-phase} and \eqref{Eq:SQL-for-frequency}.

In the case of the needle, the spin-lattice interaction leads to rapid averaging of components of $\mb{S}$ transverse to $\mb{a}$. If one had a measurement device with a sufficiently high bandwidth, in principle one could observe transverse spin projection $S_y$ with $\Delta S_y$ as described by Eq.~\eqref{Eq:spin-uncertainty}. However, a measurement device with narrower bandwidth will average over this spin projection noise. This is similar to the averaging of spin noise that occurs in some solid state experiments searching for permanent electric dipole moments \cite{Bud06}.

To estimate the quantum limit on $\Delta\phi$, we can employ the fluctuation-dissipation theorem (FDT). The physical mechanisms leading to dissipation in the form of Gilbert damping are the same ones through which $\mb{S}$ interacts with the lattice and the transverse spin components are averaged \cite{Sti01,Smi01,Eck09}. According to the FDT, in the low-frequency limit $\hbar\omega \ll k_B T$, the spectral density of transverse spin fluctuations at frequency $\omega$ is given by
\begin{align}
\prn{\delta S_y}^2 \approx \frac{V}{g^2\mu_B^2}\frac{2k_BT}{\omega}\chi''(\omega)~,
\label{Eq:FDT1}
\end{align}
where $V$ is the volume of the needle, $T$ is its temperature, and $\chi''(\omega)$ is the imaginary part of the magnetic susceptibility. Under the conditions considered here, the Landau-Lifshitz-Gilbert equation can be linearized \cite{Smi01,Bro63} to obtain the imaginary susceptibility in terms of the Gilbert damping constant $\alpha$:
\begin{align}
\chi''(\omega) \approx N\hbar\alpha\frac{g^2 \mu_B^2}{V} \frac{\omega}{\omega_0^2}~,
\label{Eq:FDT2}
\end{align}
from which we find
\begin{align}
\prn{\delta S_y}^2 \approx N\hbar \frac{2\alpha k_B T}{\omega_0^2}~.
\label{Eq:FDT3}
\end{align}
Thus for a measurement time $t$ we obtain an uncertainty in the precession angle
\begin{align}
\Delta \phi_Q \approx \frac{\delta S_y}{S_x}\frac{1}{\sqrt{t}} \approx \sqrt{\frac{2\alpha k_B T}{N\hbar\omega_0^2t}}~,
\label{Eq:FDT4}
\end{align}
and the corresponding magnetic field uncertainty is
\begin{align}
\Delta B\ts{Q} \approx \frac{\hbar}{g\mu_B}\sqrt{\frac{2\alpha k_B T}{\hbar\omega_0^2}} \frac{1}{\sqrt{Nt^3}}~.
\label{Eq:magnetometry-quantum-limit}
\end{align}
As discussed in the Supplemental Material, in order to reduce the noise from external perturbations, it is advantageous to place the needle in a cryogenic vacuum. Assuming $T \approx 0.1~{\rm K}$, $N \approx 3 \times 10^{12}$, and the values of $\alpha$ and $\omega_0$ for cobalt at $T \approx 0.1~{\rm K}$ \cite{Wat61,Bha74}, we find
\begin{align}
\Delta B\ts{Q} \approx 10^{-20} \prn{t[{\rm s}]}^{-3/2}~{\rm G}~,
\label{Eq:magnetometry-quantum-limit-numerical}
\end{align}
far below the detection-limited uncertainty $\Delta B\ts{det}$. As for the detection-limited sensitivity [Eq.~\eqref{Eq:magnetometry-SQUID-limit}], we find a $t^{-3/2}$ scaling of the measurement uncertainty, which follows from the fact that the spectral density of transverse spin fluctuations is white [Eq.~\eqref{Eq:FDT3}]. The contrast between the $t^{-1/2}$ scaling for the SQL [Eq.~\eqref{Eq:SQL-for-frequency}] and the $t^{-3/2}$ scaling of Eq.~\eqref{Eq:magnetometry-quantum-limit} is due to the fact that the thermal fluctuations of Eq.~\eqref{Eq:magnetometry-quantum-limit} come from the internal coupling between the spin and the lattice of the needle, rather than an external coupling to the environment as in the SQL. Since the needle's angular momentum is dominated by the collective spin $S$, there are only small amplitude fluctuations of the needle's angular position rather than a random walk.

The scaling of the measurement uncertainty reverts to $t^{-1/2}$ when coupling to the external environment dominates over internal fluctuations. Perturbations from the external environment, such as collisions with background gas molecules, can impart either angular momentum $dL_z$ along $\mb{B}$ or angular momentum $dL_y$ transverse to both $\mb{B}$ and $\mb{a}$. Because of the needle's gyroscopic nature, orbital angular momentum imparted by a $dL_z$ perturbation is converted into a rotation of $\mb{S}$ out of the $xy$-plane. Stochastic $dL_z$ perturbations cause a random walk of $S_z$, but as long as $\abrk{ S_z } \ll N\hbar$ measurement can continue without significant loss of sensitivity. Transverse perturbations $dL_y$ also manifest as a rotation of $\mb{S}$, but in the $xy$-plane. Therefore they are indistinguishable from transient magnetic field pulses and hence constitute a source of noise in the measurement of $\phi$. Such perturbations cause the needle to execute a random walk in $\phi$ of average step size $d\phi \approx dL_y/(N\hbar)$, and given a perturbation rate of $\Gamma_p$, the resulting spread in $\phi$ is $\Delta\phi_p \approx d\phi \sqrt{\Gamma_p t}$. Thus in a measurement time $t$ the observed precession angle noise is
\begin{align}
\Delta\phi_p \approx \frac{ dL_y }{N\hbar} \sqrt{\Gamma_pt}~.
\label{Eq:perturbation-noise}
\end{align}
Note that $\Delta\phi_p$ scales as $\sqrt{t}$ in contrast to $\Delta\phi\ts{det}$ and $\Delta\phi_Q$ which scale as $1/\sqrt{t}$. Thus after a period of measurement time, which depends on the particular experimental parameters, noise due to external perturbations dominates measurement uncertainty.

\begin{figure}
\center
\includegraphics[width=3.25 in]{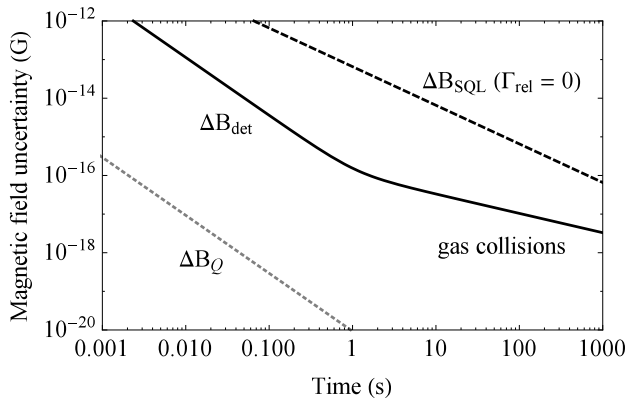}
\caption{Solid black line: magnetic field uncertainty as a function of measurement time $t$ for a precessing cobalt needle magnetometer as described in the text ($\ell = 10~{\rm \mu m}$ and $r = 1~{\rm \mu m}$, corresponding to $N \approx 3 \times 10^{12}$ spins) under conditions of cryogenic vacuum ($T \approx 0.1~{\rm K}$, background gas density $\approx 10^3~{\rm cm^{-3}}$). For $t \lesssim 1~{\rm s}$ the noise is dominated by the SQUID detector [$\Delta B\ts{det}$, Eq.~\eqref{Eq:magnetometry-SQUID-limit}], and for $t \gtrsim 1~{\rm s}$ the noise is dominated by collisions with background gas molecules. Dashed black line: standard quantum limit on magnetic field uncertainty, $\Delta B\ts{SQL}$, for $N \approx 3 \times 10^{12}$ spins with $\Gamma\ts{rel}=0$. Dotted gray line: thermally averaged quantum noise limit on magnetic field uncertainty for a precessing cobalt needle magnetometer of given dimensions at $T \approx 0.1~{\rm K}$ [$\Delta B_Q$, Eq.~\eqref{Eq:magnetometry-quantum-limit-numerical}].}
\label{Fig:sensitivity}
\end{figure}

In order to achieve the detection-limited magnetometric sensitivity described by Eq.~\eqref{Eq:magnetometry-SQUID-limit}, the needle must be well-isolated from the environment and cooled to cryogenic temperatures in order to reduce external perturbations. In the Supplemental Material, we consider noise due to collisions with residual gas molecules and black-body radiation, as well as noise from internal degrees of freedom. We find that collisions with background gas molecules become the dominant source of noise at longer measurement times. Figure~\ref{Fig:sensitivity} shows the uncertainty in the measurement of a magnetic field as a function of the measurement time for a cobalt needle of the chosen dimensions under conditions of cryogenic vacuum.

Perhaps the most daunting technical challenge for realization of a precessing needle magnetometer is the problem of suspension. Because of the stringent requirements on isolation from the environment discussed in the Supplemental Material, optical levitation \cite{Ari13} or mechanical suspension \cite{Kis06} do not appear to be viable options for an experiment aiming to reach the detection-limited measurement uncertainty $\Delta B\ts{det}$. A needle could be floated in a micro-gravity environment such as a satellite or drop tower, however it is also desirable to develop a method of ``frictionless'' suspension allowing extended measurements in an earth-bound laboratory.

One possibility is levitation of the needle above a type I superconductor. While there is non-negligible friction from flux pinning and vortices in type II superconductors, these mechanisms are suppressed in type I superconductors where levitation is based purely on flux expulsion through the Meissner effect, in principle enabling essentially frictionless levitation \cite{Bra89,Hul00,Bra85}; residual dissipation mechanisms are discussed in the Supplemental Material and Ref.~\cite{Bra85}. The drawback of such levitation is that, because of the superconductor's perfect diamagnetism, any magnetic field orthogonal to the surface that could be measured via needle precession would be counteracted by an image field, severely constraining possible applications to magnetometry.

On the other hand, exotic spin-dependent interactions are not expelled by the Meissner effect, and thus a needle magnetometer may be particulary useful for precision tests of fundamental physics. For example, the needle could be used to search for exotic spin-dependent interactions of electrons \cite{Hec13,Hun13,Bar2014,Hei05,Eck12,Kot15}. Based on the estimates shown in Fig.~\ref{Fig:sensitivity}, a measurement of needle precession averaged over $\approx 10^3~{\rm s}$ could reach a sensitivity to exotic electron-spin-dependent couplings at an energy scale of $\sim 10^{-26}~{\rm eV}$, some five orders of magnitude beyond the best constraints to date \cite{Hec13,Hun13}.

A further point of interest is that micron-scale ferromagnetic needles in the interstellar medium \cite{Dwe04} should display the predicted precession behavior in the ambient magnetic field, since typical interstellar magnetic fields in galaxies are $\sim 10^{-5}~{\rm G}$ which is on the order of $B^*$ for micron-scale needles and intergalactic magnetic fields are $\ll B^*$ but $\gtrsim 10^{-16}~{\rm G}$ \cite{Ner10}.

In conclusion, we have analyzed a micron-scale magnetometer based on measurement of the precession of a single-domain ferromagnetic needle. The needle precesses under conditions where the mechanical orbital angular momentum associated with the precession is much smaller than the intrinsic spin angular momentum of the polarized electrons in the ferromagnet. The sensitivity of a precessing needle magnetometer can surpass that of present state-of-the-art magnetometers by several orders of magnitude.

The authors are sincerely grateful to Eugene Commins, Max Zolotorev, Erwin Hahn, Marcis Auzinsh, Alex Zettl, Michael Crommie, Oleg Sushkov, Victor Flambaum, Mikhail Kozlov, John Clarke, Lutz Trahms, Kathryn A. Moler, Szymon Pustelny, Holger M\"uller, Jennie Guzman, Jason Singley, Erik Helgren, Nathan Leefer, Hunter Richards, Surjeet Rajendran, Amir Yacobi, Mikhail Lukin, and Asimina Arvanitaki for enlightening discussions.  This work was supported by the Heising-Simons Foundation, the Simons Foundation, and the National Science Foundation under grant PHY-1307507. The authors are grateful to the Mainz Institute for Theoretical Physics (MITP) for its hospitality and its partial support during the completion of this work.


\begin{thebibliography}{99}

\bibitem{Bra75}
V. B. Braginsky and Y. I. Vorontsov, Sov. Phys. Usp. {\textbf{17}}, 644 (1975).


\bibitem{Win92}
D. J. Wineland, J. J. Bollinger, W.M. Itano, F. L. Moore, and D. J. Heinzen, Phys. Rev. A {\textbf{46}}, R6797 (1992).

\bibitem{Hue97}
S. F. Huelga, C. Macchiavello, T. Pellizzari, A. K. Ekert, M. B. Plenio, and J. I. Cirac, Phys. Rev. Lett. {\textbf{79}}, 3865 (1997).

\bibitem{Auz04}
M. Auzinsh, D. Budker, D. F. Kimball, S. M. Rochester, J. E. Stalnaker, A. O. Sushkov, and V. V. Yashchuk, Phys. Rev. Lett. {\textbf{93}}, 173002 (2004).

\bibitem{Sha10}
V. Shah, G. Vasilakis, and M. V. Romalis, Phys. Rev. Lett. {\textbf{104}}, 013601 (2010).

\bibitem{Was10}
W. Wasilewski, K. Jensen, H. Krauter, J. J. Renema, M. V. Balabas, and E. S. Polzik, Phys. Rev. Lett. {\textbf{104}}, 133601 (2010).

\bibitem{Sew12}
R. J. Sewell, M. Koschorreck, M. Napolitano, B. Dubost, N. Behbood, and M. W. Mitchell, Phys. Rev. Lett. {\textbf{109}}, 253605 (2012).

\bibitem{Wol12}
F. Wolfgramm, C. Vitelli, F. A. Beduini, N. Godbout, and M. W. Mitchell, Nature Photon. {\textbf{7}}, 28 (2012).



\bibitem{Bud07}
D. Budker and M. V. Romalis, Nature Phys. {\textbf{3}}, 227 (2007).

\bibitem{Bud13book}
D. Budker and D. F. Jackson Kimball, eds., {\it{Optical Magnetometry}} (Cambridge University Press, Cambridge, 2013).

\bibitem{Bud02review}
D. Budker, W. Gawlik, D. F. Kimball, S. M. Rochester, V. V. Yashchuk, and A. Weiss, Rev. Mod. Phys. {\textbf{74}}, 1153 (2002).




\bibitem{Dan10}
H. B. Dang, A. C. Maloof, and M.V. Romalis, Appl. Phys. Lett. {\textbf{97}}, 151110 (2010).

\bibitem{She13}
D. Sheng, S. Li, N. Dural, and M.V. Romalis, Phys. Rev. Lett. {\textbf{110}}, 160802 (2013).

\bibitem{All02}
J. C. Allred, R. N. Lyman, T. W. Kornack, and M. V. Romalis, Phys. Rev. Lett. {\textbf{89}}, 130801 (2002).

\bibitem{Kom03}
I. K. Kominis, T. W. Kornack, J. C. Allred, and M. V. Romalis, Nature {\textbf{422}}, 596 (2003).

\bibitem{Sav05}
I. M. Savukov and M. V. Romalis, Phys. Rev. A {\textbf{71}}, 023405 (2005).





\bibitem{Gau1833}
C. F. Gauss, {\it{Intensitas vis magneticae terrestris ad mensuram absolutam revocata.}} (Gottingen, 1833).

\bibitem{Bud08book}
D. Budker, D. F. Kimball, and D. P. DeMille, {\it{Atomic Physics: an exploration through problems and solutions}} 2nd edition (Oxford University Press, Oxford, 2008).



\bibitem{Aha88}
A. Aharoni, J. Appl. Phys. {\textbf{63}}, 5879 (1988).

\bibitem{Sey01}
E. Seynaeve, G. Rens, A. V. Volodin, K. Temst, C. Van Haesendonck, and Y. Bruynseraede,  J. Appl. Phys. {\textbf{89}}, 531 (2001).

\bibitem{Led94}
M. Lederman, S. Schultz, and M. Ozaki, Phys. Rev. Lett. {\textbf{73}}, 1986 (1994).

\bibitem{Chi97}
S. Chikazumi, {\it Physics of Ferromagnetism} 2nd edition (Oxford University Press, Oxford, 1997).



\bibitem{Lan35}
L. D. Landau and L. M. Lifshitz, Phys. Z. Sowjetunion {\textbf{8}}, 153 (1935).

\bibitem{Gil04}
T. L. Gilbert, IEEE Trans. Magn. {\textbf{40}}, 3443 (2004).

\bibitem{Xia05}
J. Xiao, A. Zangwill, and M. D. Stiles, Phys. Rev. B {\textbf{72}}, 014446 (2005).

\bibitem{Bra08}
A. Brataas, Y. Tserkovnyak, and G. E. W. Bauer, Phys. Rev. Lett. {\textbf{101}}, 037207 (2008).

\bibitem{Bha74}
S. M. Bhagat and P. Lubitz, Phys. Rev. B {\textbf{10}}, 179 (1974).

\bibitem{Kat00}
J. A. Katine, F. J. Albert, R. A. Buhrman, E. B. Myers, and D. C. Ralph, Phys. Rev. Lett. {\textbf{84}}, 3149 (2000).

\bibitem{Sch95}
F. Schreiber, J. Pflaum, Z. Frait, T. M\"uhge, and J. Pelzl, Solid State Commun. {\textbf{93}}, 965 (1995).

\bibitem{Tse02}
Y. Tserkovnyak, A. Brataas, and G. E. W. Bauer, Phys. Rev. Lett. {\textbf{88}}, 117601 (2002).

\bibitem{Wat61}
R. E. Watson and A. J. Freeman Phys. Rev. {\textbf{123}}, 2027 (1961).

\bibitem{Gil07}
K. Gilmore, Y. U. Idzerda, and M. D. Stiles, Phys. Rev. Lett. {\textbf{99}}, 027204 (2007).


\bibitem{Mos58}
R. L. M\"ossbauer, Z. Physik A {\textbf{151}}, 124 (1958).

\bibitem{Bar14}
S. J. Barnett, Phys. Rev. {\textbf{6}}, 239 (1915).

\bibitem{Bar35}
S. J. Barnett, Rev. Mod. Phys. {\textbf{7}}, 129 (1935).

\bibitem{Ein15}
A. Einstein and W. J. de Haas, Verhandl. Deut. Phsik. Ges. {\textbf{17}}, 152 (1915); {\it{ibid}}. {\textbf{18}}, 173 (1916).

\bibitem{Alz67}
G. Alzetta, E. Arimondo, C. Ascoli, and A. Gozzini, Il Nuovo Cim. {\textbf{52}}, 392 (1967).

\bibitem{Los16}
J. E. Losby and M. R. Freeman, arXiv:1601.00674 (2016).


\bibitem{Pan04}
M. Pannetier, C. Fermon, G. Le Goff, J. Simola, and Emma Kerr, Science {\textbf{304}}, 1648 (2004).

\bibitem{Aws88}
D. D. Awschalom, J. R. Rozen, M. B. Ketchen, W. J. Gallagher, A. W. Kleinsasser, R. L. Sandstrom, and B. Bumble,  Appl. Phys. Lett. {\textbf{53}}, 2108 (1988).

\bibitem{Muc01}
Michael M\"uck, J. B. Kycia, and John Clarke, Appl. Phys. Lett. {\textbf{78}}, 967 (2001).

\bibitem{Use11}
O. Usenko, A. Vinante, G. Wijts, and T. H. Oosterkamp, Appl. Phys. Lett. {\textbf{98}}, 133105 (2011).

\bibitem{Hub08}
W. Gardner, Sean T. Halloran, Erik A. Lucero, and Kathryn A. Moler, Rev. Sci. Inst. {\textbf{79}}, 053704 (2008).

\bibitem{Cla04v1}
J. Clarke and A. I. Braginski, eds., {\it{The SQUID Handbook Vol. 1}} (Wiley, Weinheim, 2004).

\bibitem{Tes90}
C. D. Tesche, Phys. Rev. Lett. {\textbf{64}}, 2358 (1990).

\bibitem{Led05}
M. Ledbetter, {\it{Progress Toward a Search for a Permanent Electric Dipole Moment in Liquid $^{129}$Xe}} (Ph.D. dissertation, Princeton University, 2005).

\bibitem{Poo10}
M. Poot, S. Etaki, I. Mahboob, K. Onomitsu, H. Yamaguchi, Ya. M. Blanter, and H. S. J. van der Zant, Phys. Rev. Lett. {\textbf{105}}, 207203 (2010).

\bibitem{Hat11}
M. Hatridge, R. Vijay, D. H. Slichter, J. Clarke, and I. Siddiqi, Phys. Rev. B {\textbf{83}}, 134501 (2011).



\bibitem{Bud06}
D. Budker, S. K. Lamoreaux, A. O. Sushkov, and O. P. Sushkov, Phys. Rev. A {\textbf{73}}, 022107 (2006).



\bibitem{Sti01}
B. C. Stipe, H. J. Mamin, C. S. Yannoni, T. D. Stowe, T. W. Kenny, and D. Rugar, Phys. Rev. Lett. {\textbf{87}}, 277602 (2001).

\bibitem{Smi01}
N. Smith and P. Arnett, Appl. Phys. Lett. {\textbf{78}}, 1448 (2001).

\bibitem{Eck09}
S. Eckel, A. O. Sushkov, and S. K. Lamoreaux, Phys. Rev. B {\textbf{79}}, 014422 (2009).

\bibitem{Bro63}
W. F. Brown, Phys. Rev. {\textbf{130}}, 1677 (1963).




\bibitem{Ari13}
Y. Arita, M. Mazilu, and K. Dholakia, Nature Commun. {\textbf{4}}, 2374 (2013).

\bibitem{Kis06}
A. Kis, K. Jensen, S. Aloni, W. Mickelson, and A. Zettl Phys. Rev. Lett. {\textbf{97}}, 025501 (2006).



\bibitem{Bra89}
E. H. Brandt, Science {\textbf{243}}, 349 (1989).

\bibitem{Hul00}
J. R. Hull, Supercond. Sci. Technol. {\textbf{13}}, R1 (2000).

\bibitem{Bra85}
V. B. Braginsky, V. P. Mitrofanov, and V. I. Panov, {\it{ Systems with Small Dissipation }} (University of Chicago Press, Chicago, 1985).



\bibitem{Hec13}
B. R. Heckel, W. A. Terrano, and E. G. Adelberger, Phys. Rev. Lett. {\textbf{111}}, 151802 (2013).

\bibitem{Hun13}
L. Hunter, J. Gordon, S. Peck, D. Ang, and J.-F. Lin, Science {\textbf{339}}, 928 (2013).

\bibitem{Bar2014}
J. Baron, et al. Science {\textbf{343}}, 269 (2014).


\bibitem{Hei05}
B. J. Heidenreich, et al. Phys. Rev. Lett. {\textbf{95}}, 253004 (2005).



\bibitem{Eck12}
S. Eckel, A. O. Sushkov, and S. K. Lamoreaux, Phys. Rev. Lett. {\textbf{109}}, 193003 (2012).

\bibitem{Kot15}
S. Kotler, R. Ozeri, and D. F. Jackson Kimball, Phys. Rev. Lett. {\textbf{115}}, 081801 (2015).


\bibitem{Dwe04}
Eli Dwek, Astro. Phys. J. {\textbf{607}}, 848 (2004).

\bibitem{Ner10}
A. Neronov and I. Vovk, Science {\textbf{328}}, 73 (2010).




\end{thebibliography}
\end{document}


\title{Supplemental Material for a Precessing Ferromagnetic Needle Magnetometer} 

\date{\today}



\author{Derek F. Jackson Kimball}
\affiliation{Department of Physics, California State University -- East Bay, Hayward, California 94542-3084, USA}

\author{Alexander O. Sushkov}
\affiliation{Department of Physics, Boston University, Boston, Massachusetts 02215, USA}

\author{Dmitry Budker}
\affiliation{Helmholtz Institute Mainz, Johannes Gutenberg University, 55099 Mainz, Germany}
\affiliation{Department of Physics, University of California at Berkeley, Berkeley, California 94720-7300, USA}
\affiliation{Nuclear Science Division, Lawrence Berkeley National Laboratory, Berkeley, California 94720, USA}

\begin{abstract}
This note discusses the thermal noise of a precessing ferromagnetic needle magnetometer from both internal degrees of freedom and external perturbations, as well as other issues related to practical realization of a precessing ferromagnetic needle magnetometer.
\end{abstract}



\maketitle

\section{Internal degrees of freedom}

\subsection{Magnons and phonons}

Gyroscopic stability is a key feature of the ferromagnetic needle in the regime where its dynamics are dominated by its intrinsic spin angular momentum (the regime where $S \gg L$). Due to conservation of angular momentum, the total angular momentum $\mb{J} = \mb{L} + \mb{S} \approx \mb{S}$ is fixed in the absence of external perturbations. Assuming no external perturbations, the only way for thermal fluctuations of $\mb{S}$ to occur is through a spin-lattice interaction causing corresponding fluctuations of $\mb{L}$ that leave $\mb{J}$ unchanged. Such thermal fluctuations are precisely the noise described in the main text in terms of the fluctuation-dissipation theorem: the same spin-lattice interactions that lead to Gilbert damping are those that couple $\mb{S}$ to $\mb{L}$ and lead to the spin noise described in Eqs. (11)-(13) of the main text. These spin fluctuations encompass the noise related to magnons \cite{Kit04} and phonons (which can carry angular momentum \cite{Zha14,Gar15}).

The combination of crystalline and shape anisotropy for a $\sim 1~{\rm \mu m} \times 10~{\rm \mu m}$ Co needle creates a significant energy gap (the magnon gap) on the order of 1 K between the ground and first excited states for the collective spin \cite{Sey01,Sin60,Hul86,Aki88}. This means that if the needle is cooled to a temperature $\ll 1~{\rm K}$, higher-order magnon modes can be effectively frozen out and we only need to consider the lowest-order uniform precession mode (where the entire collective spin $\mb{S}$ rotates with respect to the anisotropy axis) as we do in the main text where we employ the macrospin model.

\subsection{Thermal currents}

\begin{figure}
\center
\includegraphics[width=3.25 in]{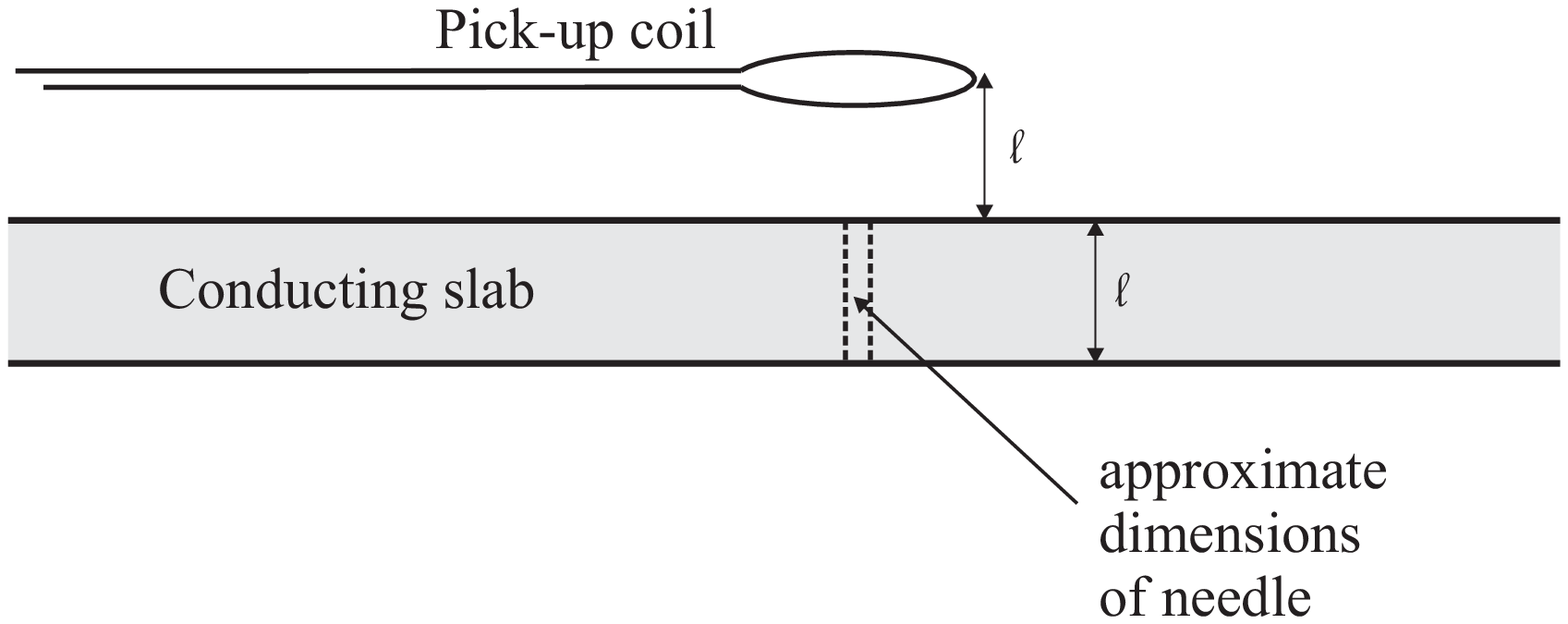}
\caption{Geometry for estimate of noise due thermal currents. The analytical expression in Eq.~\eqref{Eq:thermal-currents} is for a thin conducting slab of thickness $\ell$, and thus the noise from a conducting needle of radius $r \approx \ell/10$ should be considerably less.}
\label{Fig:sensitivity}
\end{figure}

Johnson-Nyquist noise from thermal currents in the electrically conductive cobalt needle can generate magnetic field noise in the SQUID pick-up loop \cite{Var84,Lam99,Sti01,Lee08}. Although exact calculations for most geometries are complicated and generally require numerical integration, to estimate an upper limit on such noise we assume the needle can be treated as a thin conducting slab of thickness $\approx \ell$ (the length of the needle) for which an analytical expression for the magnetic noise spectral density is available \cite{Var84}:
\begin{align}
\prn{ \delta B^2 } \approx \frac{\pi}{4} \frac{k_B T}{c^2\rho\ell}~,
\label{Eq:thermal-currents}
\end{align}
where $\ell \approx 10~{\rm \mu m}$ is the distance from the needle to the SQUID pick-up loop and $\rho$ is the resistivity of cobalt (for $T \lesssim {\rm 1~K}$, $\rho \approx 10^{-7}~\Omega \cdot {\rm cm}$ \cite{Hal68}). According to this estimate, the fluctuating field amplitude at the SQUID pick-up loop due to thermal currents should be
\begin{align}
\delta B \ll 10^{-8}~\frac{\rm G}{\sqrt{\rm Hz}}~.
\end{align}
Multiplying this field value by the area of the pick-up loop we find that the flux noise $\delta \Phi \ll 10^{-13}~{\rm G \cdot cm^2 / \sqrt{ Hz }}$, which implies that noise due to thermal currents should be smaller than the sensitivity of low-temperature SQUIDs to flux changes. Thus thermal currents do not affect our estimate for the detection-limited sensitivity $\Delta B\ts{det}$. Furthermore, in the event this becomes a limiting factor for a precessing needle magnetometer, one could consider the use of nonconducting ferrimagnets as the needle material which can have resistivities many orders of magnitude larger than that of cobalt.

\section{External perturbations}

\subsection{Collisions with residual gas molecules}

Collisions with residual gas molecules (assumed to be He) impart angular momentum to the needle with average magnitude
\begin{align}
dL\ts{col} \approx \frac{m v \ell}{16} \approx 10^3 \hbar~,
\label{Eq:collisions-dL}
\end{align}
where $m$ is the mass of He, $v \approx 3 \times 10^{3}~{\rm cm/s}$ is the average thermal velocity for $T \approx 0.1~{\rm K}$, and the factor of 1/16 arises from averaging over the angle and location of impact. The collision rate is
\begin{align}
\Gamma\ts{col} \approx \frac{nAv}{4}~,
\label{Eq:collisions-rate}
\end{align}
where $A \approx r\ell \approx 10^{-7}~{\rm cm^2}$ is the relevant cross sectional area of the needle and $n$ is the residual gas density. Substituting the expressions \eqref{Eq:collisions-dL} and \eqref{Eq:collisions-rate} into Eq.~(15) from the main text yields
\begin{align}
\Delta \phi\ts{col} \approx \frac{m}{32 N \hbar} \sqrt{ n r \ell^3 v^3 t}~,
\end{align}
which corresponds to a magnetic field uncertainty of
\begin{align}
\Delta B\ts{col} \approx \frac{\hbar}{g\mu_B}\frac{m}{32 N \hbar} \sqrt{ \frac{n r \ell^3 v^3}{t} }~.
\end{align}
Requiring $\Delta B\ts{col}$ to be less than the detection-limited measurement sensitivity $\Delta B\ts{det}$ described by Eq.~(6) from the main text for measurement times $t \lesssim 1~{\rm s}$ constrains $n \lesssim 10^3~{\rm atoms/cm^{3}}$. Note that on average, according to Eq.~\eqref{Eq:collisions-rate}, under these conditions collisions with gas molecules happen about once a second, so to reach the detection-limited sensitivity essentially no collisions can occur during the measurement time. Such ultralow residual gas densities have been achieved, for example, in ion trapping experiments \cite{Gab90,Die98} under cryogenic vacuum conditions. For measurement times $t \gtrsim 1~{\rm s}$, noise due to collisions with background gas molecules dominates (see Fig.~2 in the main text).

The scaling of $\Delta B\ts{col}$ with the size of the needle can be estimated by assuming a fixed aspect ratio, in which case both $\ell$ and $r$ are proportional to $N^{1/3}$. In this case, $\Delta B\ts{col} \propto 1/N^{1/3}$, so, as in the case of detection- and quantum-limited uncertainty, larger ferromagnetic needles can achieve greater sensitivity in principle.

Related effects make mechanical suspension of a needle by a fiber while retaining the exceptional magnetometric sensitivity difficult. If the fiber exerts a torque on the needle, then it acts as a conduit of angular momentum transfer between the needle and the environment and causes stochastic fluctuations far in excess of those produced by collisions with background gas molecules.

\subsection{Black-body radiation}

Photons from black-body radiation are another source of external perturbations. According to the Stefan-Boltzmann law, the number of photons emitted by the needle per second is
\begin{align}
\Gamma\ts{BB} = \frac{4 \zeta(3) \varepsilon}{c^2 h^3} k_B^3 T^3 2\pi A~,
\label{Eq:Stefan-Boltzmann}
\end{align}
where $\zeta(3) \approx 1.2$ is the Riemann zeta function with argument 3, $\varepsilon$ is the emissivity, and $2\pi A \approx 2\pi r \ell \approx 6 \times 10^{-7}~{\rm cm^2}$ is the surface area of the needle. At $T \approx 0.1~{\rm K}$, Eq.~\eqref{Eq:Stefan-Boltzmann} yields $\Gamma\ts{BB} \approx 100\varepsilon~{\rm photons/s}$. Since the characteristic wavelength of black-body radiation is $\lambda\ts{BB} \approx 3~{\rm mm}$ or roughly $300$ times the dimensions of the needle, $\varepsilon$ should be suppressed by over an order of magnitude \cite{Wut13}. Furthermore, in this regime, the needle absorbs and emits radiation as a point-like dipole \cite{Cha09,Gol12}, so the coupling of the photon momentum to the macroscopic rotational motion of the needle is negligible. However, random polarization of the black-body photons can generate longitudinal and transverse perturbations of the needle of magnitude $dL\ts{BB} \approx \hbar$ per photon. The effect of such stochastic kicks from blackbody photons on the measurement sensitivity can be analyzed in the same way as was done in the previous section for the effect of collisions with gas molecules. Yet the noise from black-body radiation estimated based on Eq.~(15) from the main text is far below that due to collisions with residual gas molecules. On the other hand, stochastic noise from scattered photons rules out the use of optical suspension of the needle.

\section{Levitation above a Type I superconductor}

One solution to the problem of frictionless suspension of the needle proposed in the main text is to levitate the needle above a type I superconductor using the Meissner effect. Below their critical temperature, superconductors completely lose their resistance to direct current; on the other hand, there remains a small resistance to alternating currents due to nonzero surface impedance within the London penetration depth \cite{Bra85}. However, for type I superconductors this effect becomes vanishingly small for low temperatures (well below the critical temperature) and low frequencies (well below the frequency corresponding to the superconducting energy gap) \cite{Bra85}, which are the conditions for operation of the precessing needle magnetometer. Under these conditions, residual resistance remains due to various material-dependent electromagnetic loss mechanisms: surface smoothness and contamination are important factors, as well as magnetic flux trapped during the transition to the superconducting state that creates regions of normal conductivity. High-Q superconducting microwave cavities ($Q \gtrsim 10^{10}$ \cite{Pie73}) and magnetically levitated mechanical oscillators ($Q \gtrsim 10^{8}$ \cite{Fer79,Bla79}) have been constructed. One of the dominant frictional effects in these cases comes from eddy currents induced in non-superconducting metallic material \cite{Bra88} present in the experiments. There should be negligible loss due to eddy currents in the needle itself, since, based on the symmetry of the superconductor with respect to rotation of the needle, there is no change of the magnetic field in the needle's frame and consequently no induced eddy currents in the needle. Ultimately, the friction in a system based on levitating a ferromagnetic needle above a type I superconductor will likely be determined by incomplete flux exclusion due to defects and impurities in the superconductor \cite{Ben99}.

\section{Additional considerations for practical realization}

Finally, we point out several other factors to be considered in regards to practical realization of a precessing ferromagnetic needle magnetometer.

In the main text, we choose the size, shape, and material of the ferromagnetic needle such that the lowest energy state is a single domain \cite{Aha88,Sey01,Aha01}. However, due to hysteresis, real ferromagnetic particles can become stuck in higher-energy multi-domain states, and there can also be magnetic microstructure arising from impurities and edge domains. Therefore in order to obtain a single-domain needle, care must be taken in selecting the sample. Multi-domain needles can have additional noise due to thermal fluctuations of magnetic domain walls.

Magnetic field gradients must also be controlled so as not to exert an uncontrolled force on the needle moving it out of the detection region during a measurement. Assuming the needle is freely floating (for example, in a microgravity environment) and begins at rest, and requiring that needle's displacement is less than its radius $r$ during the measurement time $t$, we find that, for example, the gradient along $x$ is constrained by
\begin{align}
\left| \pdbyd{B_x}{x} \right| < \frac{2rm_a}{\mu_Bt^2}~.
\end{align}
For a measurement time of $t \approx 1~{\rm s}$, $\left| \partial B_x/\partial x \right| < 2~{\rm \mu G/cm}$. Gradients have been controlled to smaller than this level in many experiments \cite{Pus06}.

The above discussion again emphasizes an important technical issue: to achieve the detection-limited sensitivity described in the main text, any method to prepare, trap, or support the needle will have to be carefully designed to minimize coupling to the environment. We plan to explore these and other issues related to experimental realization of a precessing ferromagnetic needle magnetometer in future work.